# Cryptanalyze and design strong S-Box using 2D chaotic map and apply to irreversible key expansion

Hongjun Liu [a, *], Xingyuan Wang [b]

[a] School of Mathematical Sciences, University of Jinan, Jinan, Shandong, 250022, China

[b] School of Information Science and Technology, Dalian Maritime University, Dalian 116026, China

**Abstract:** Cryptanalysis result of key expansion algorithms in AES and SM4 revealed that, (1) there exist weaknesses in their S-Boxes, and (2) the round key expansion algorithm is reversible, i.e., the initial key can be recovered from any round key, which may be an exploitable weakness by attacker. To solve these problems, first we constructed a non-degenerate 2D exponential hyper chaotic map (2D-ECM), derived the recursion formula to calculate the number of S-Boxes that satisfied three conditions, and designed a strong S-Box construction algorithm without weakness. Then based on 2D-ECM and S-Box, we designed an irreversible key expansion algorithm, to transform the initial key into independent round keys, to make the initial key can not be recovered from any round key. Security and statistical analysis demonstrated the flexible and effectiveness of the proposed irreversible key expansion algorithm.

# 1. Introduction

Key expansion plays an important role in block ciphers, such as AES and SM4, it can transform a $n$-bit initial key into $r\times n$-bit round keys, whose nonlinearity and complexity rely on the round constant and S-Box. The security of a block cipher depends a great deal on the round keys during the encryption process, and improper key expansion may lead to malicious actors being able to predict the initial key [1].

Generally, a cryptographical S-Box can satisfy six criteria [2], furthermore, a strong S-Box should further satisfy the following three conditions: (i) without fixed point, (ii) without reverse fixed point, and (iii) without short period rings. However, S-Box in AES [3] only satisfies conditions (i) and (ii), S-Box in SM4 [4] only satisfy condition (ii), and for two S-Boxes in ZUC, $S_0$ only satisfies condition (i), and $S_1$ satisfies conditions (i) and (ii).

An 8×8 S-Box ∈ [0, 255] belongs to the subset of all permutations of 256!, for any S-Box that only satisfies three conditions above, it is not sure that it could satisfy the six criteria. The number of S-Boxes that satisfies six criteria is difficult to pinpoint, because constructing an S-Box that is resistant to cryptanalysis, belongs to the Non-deterministic Polynomial Complete (NPC) problem, however, we can calculate the total number of S-Boxes exactly that satisfies three conditions above, to estimate the order of magnitudes.

Arising from the limited precision of computer, the iteration process of any chaotic map is irreversible [5], which can be an accessible characteristic to generate round keys. We can take a chaotic map with ergodicity as a transverter, to transform the initial key into round keys through iterations with precision lossy, to ensure each round key independent of each other, and can not derive the initial key from any round key.

------------------------------------

[*] Corresponding author. E-mail address: sms_liuhj@ujn.edu.cn (Liu H J).

In this paper, we will reveal the weaknesses in S-Box, calculate the number of S-Boxes that satisfies three conditions, then design an enhanced key expansion algorithm. First we need to design a 2D-ECM with ergodicity, then construct a strong S-Box without fixed-point, reverse fixed-point, or short period ring, and based on them to design an irreversible key expansion algorithm. Finally, use security and statistical analysis to evaluate the effectiveness of the proposed key expansion algorithm.

The remainder parts are organized as follows: Section 2 is to reveal the weaknesses in S-Box. Section 3 is to analyze the key expansion algorithm in AES and SM4. Section 4 proposes a 2D-ECM with dynamic analysis. Section 5 is to construct a strong S-Box based on 2D-ECM. Section 6 applies 2D-ECM and S-Box to key expansion algorithm, and Section 7 reaches a conclusion.

# 2. Cryptanalysis of S-Box

## 2.1 Six criteria

Generally, a cryptographic S-Box should satisfy the following six criteria.

(1) Bijective. The ideal value of a $n\times n$ S-Box is $2^{n-1}$ [6, 7], hence, if $n$=8, the ideal value is 128.

(2) Nonlinearity. For an 8×8 S-Box, Picek et al. [8] considered that the best possible nonlinearity is 118. Isa et al. [9] considered that an 8×8 S-Box must have a nonlinearity≥100 to be considered cryptographically strong, and Dragan [10] considered that an S-Box having a nonlinearity≥106 to be high quality. Miroslav [11] designed a chaos-based S-Box with the nonolinearity of 114.

(3) Strict avalanche criterion (SAC). An S-Box achieving the ideal value 0.5 [12] indicates that every output bit is completely uncorrelated to any input bit.

(4) Output bit independence criterion (BIC). The BIC of an S-Box should be close to the ideal value 0.5 [12], which indicates that the output bits are pair-wise uncorrelated.

(5) Differential approximation probability (DAP). The value of DAP should be close to zero, which indicates that an S-Box is highly immune to differential cryptanalysis attack [13].

(6) Linear approximation probability (LAP). The value of LAP should be close to zero, which indicates that the output bits of an S-Box are uncorrelated to the input bits [13].

## 2.2 Number of S-Boxes that satisfies three conditions

### 2.2.1 Number of S-Boxes that satisfies condition (i)

Which is known as the Bernoulli-Euler problem of the misaddressed letters [14]. Suppose the number of all the permutations of $n$ elements that satisfies condition (i) is $D_1(n)$, according to the staggered formula, the recursion formula can be expressed using Eq. (1).

$$D_1(n) = (n-1)(D_1(n-1) + D_1(n-2)),\ \ n \geq 3\,, \tag{1}$$

where $D_1(1)=0$, $D_1(2)=1$. Furthermore, the general term formula of Eq. (1) can be expressed as Eq. (2).

$$D_1(n)=[\,n!/e+0.5\,], \tag{2}$$

where $e$ stands for the nature constant. For an 8×8 S-Box$\in$[0, 255], $n$=256, $D_1(256) \approx 3.1557\times 10^{506}$.

### 2.2.2 Number of S-Boxes that satisfies conditions (i) and (ii)

Suppose $D_2(n)$ is the number of all permutations of $n$ elements that satisfies conditions (i) and (ii), through exhaustive and derivation, we have obtained that $D_2(1)=0$, $D_2(2)=0$, $D_2(3)=0$, $D_2(4)=4$, $D_2(5)=16$, $D_2(6)=80$, $D_2(7)=672$, $D_2(8)=4752$, and $D_2(9)=48768$, and based on them, we can derive a recursion formula, which can be expressed using Eq. (3).

$$D_2(n)=\begin{cases}(n-1)(D_2(n-1)+2D_2(n-2)), \text{ if odd } n\geq 5\\(n-2)(D_2(n-1)+D_2(n-2)+2D_2(n-3)+2D_2(n-4)), \text{ if even } n\geq 6\end{cases}. \quad (3)$$

For an 8×8 S-Box ∈ [0, 255], $n$ =256, the number of $D_2(256)\approx 1.1564\times 10^{506}$.

### 2.2.3 Number of S-Boxes that satisfies conditions (i), (ii) and (iii)

Suppose $D_3(n)$ is the number of all permutations of $n$ elements that satisfies conditions (i), (ii) and (iii), through exhaustion and derivation, we have obtained that $D_3(1)=0$, $D_3(2)=0$, $D_3(3)=0$, $D_3(4)=2$, $D_3(5)=8$, $D_3(6)=32$, $D_3(7)=240$, $D_3(8)=1488$, and $D_3(9)=13824$. Based on them, we can derive a recursion formula that can be expressed using Eq. (4).

$$D_3(n)=\begin{cases}(n-1)(D_3(n-1)+D_3(n-2)), \text{if odd } n\geq 3\\(n-2)(D_3(n-1)+D_3(n-3)), \text{if even } n\geq 4\end{cases}. \quad (4)$$

For an 8×8 S-Box ∈ [0, 255], $n$ =256, the number of $D_3(256)\approx 1.2255\times 10^{504}$, and of all the permutations $256!\approx 8.5782\times 10^{506}$, $D_3(256)$ merely accounts for about 1.4286‰, here we can define any S-Box that satisfies six criteria and three conditions as a strong S-Box. However, except for exhaustion, to define a function and walk through all the S-Boxes in $D_3(256)$ is extremely difficult.

Table 1 Periodic rings of S-Box in AES and SM4

| | Length of periodic rings | Short periodic rings | Length |
|---|---|---|---|
| S-Box in AES | 2, 27, 59, 81, 87 | 73→8F→73 | 2 |
| S-Box in Skipjack | 2, 10, 45, 68 | 78→E1→78 | 2 |
| S-Box in Whirlpool | 3, 6, 32, 91 | 95→3A→E7→95 | 3 |
| | | D9→96→DE→BC→53→60→D9 | 6 |
| S-Box in SM4 | 1, 2, 3, 6, 9, 24, 35, 56, 120 | AB (Fixed point) | 1 |
| | | B4→DE→B4 | 2 |
| | | D8→2D→CF→D8 | 3 |
| | | 65→58→F8→79→36→E8→65 | 6 |
| | | AE→4E→4F→A8→C0→8D→61→24→91→AE | 9 |

# 3 Cryptanalysis of key expansion in AES and SM4

After careful analysis, we have found that there exist weaknesses in the key expansion algorithms of AES and SM4. For example, the weaknesses of key expansion in AES include: (1) there exists a very short period rings in S-Box, i.e., SBox(‘B4’)=‘DE’, and SBox(‘DE’)=‘B4’, as shown in Table 1; (2) there exists linear transformation in the generation of round constants, which are expressed as Rcon[*i*]=(RC[*i*], 0, 0, 0), where RC[1]=1, RC[*i*]=2×RC[*i*-1], although the multiplication is defined over GF($2^8$), from Table 2 we can infer that except for the initial value RC[1] and RC[10], the rest values of RC[*i*] are generated from linear transformation actually; (3) the key expansion process is reversible, i.e., the initial key can be recovered from any round key $RK_i$, $i \in [1,n]$, $n$=10, 12 or 14, as shown in Fig. 1.

Table 2 The round constant RC[*j*] in AES

| *j* | 1 | 2 | 3 | 4 | 5 | 6 | 7 | 8 | 9 | 10 |
|---|---|---|---|---|---|---|---|---|---|---|
| RC[*j*] | 01 | 02 | 04 | 08 | 10 | 20 | 40 | 80 | 1B | 36 |

Compare with AES, the round constants in key expansion algorithm of SM4 [4] is enhanced significantly, but its S-Box still have several weaknesses: (1) there exists one fixed point, i.e., SBox('AB')='AB'; (2) there exist four very short periodic rings, as shown in Table 1, which may result in finite loop substitution; and (3), just like AES, the key expansion process is reversible, hence, the initial key can be recovered from any round key $RK_i$, $i \in [1,n]$, $n$=32, as shown in Fig. 1.

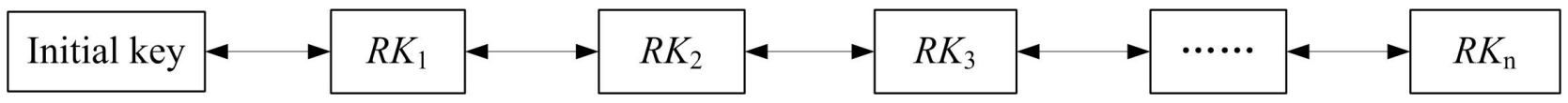

Fig. 1 Reversible key expansion process in AES and SM4

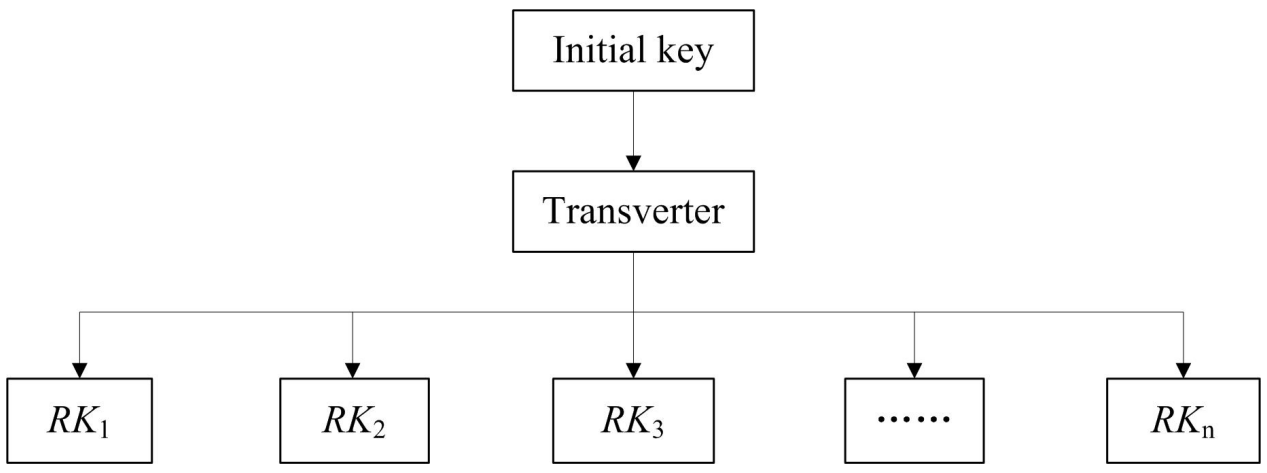

Fig. 2 Irreversible key expansion process

Hence, a reversible key expansion algorithm cannot ensure the security of initial key, we need to design an irreversible key expansion algorithm, to make the round keys independent of each other, as shown in Fig. 2, which can ensure a higher security for block cipher.

# 4. 2D-ECM construction and dynamics analysis

## 4.1 Two 1D seed maps

First, let's review two classic 1D discrete time chaotic maps: Logistic map and Quadratic map, they would be served as seed maps to construct a 2D-ECM.

The difference equation of Logistic map proposed by Robert May [15] can be expressed using Eq. (5).

$$x_{i+1} = \mu(x_i - x_i^2), \tag{5}$$

where $x_i \in [0,1]$ refers to the state variable, and $\mu \in [0,4]$ is the control parameter.

The equation of Quadratic map [16] is similar to Eq. (5), and can be expressed using Eq. (6).

$$x_{i+1} = \gamma - x_i^2, \tag{6}$$

where $x_i \in [-2, 2]$ refers to the state variable, and $\gamma \in [0, 2]$ is the control parameter.

From the bifurcation diagrams and phase diagrams shown in Fig. 3 and Fig. 4, we can infer that these 1D seed maps have simple structure but complex dynamics behavior within their finite parameter range, and there exists strong correlation between two adjacent state points. However, they can be served as seed maps to construct a non-degenerate 2D-ECM with ergodicity within a larger range of control parameter.

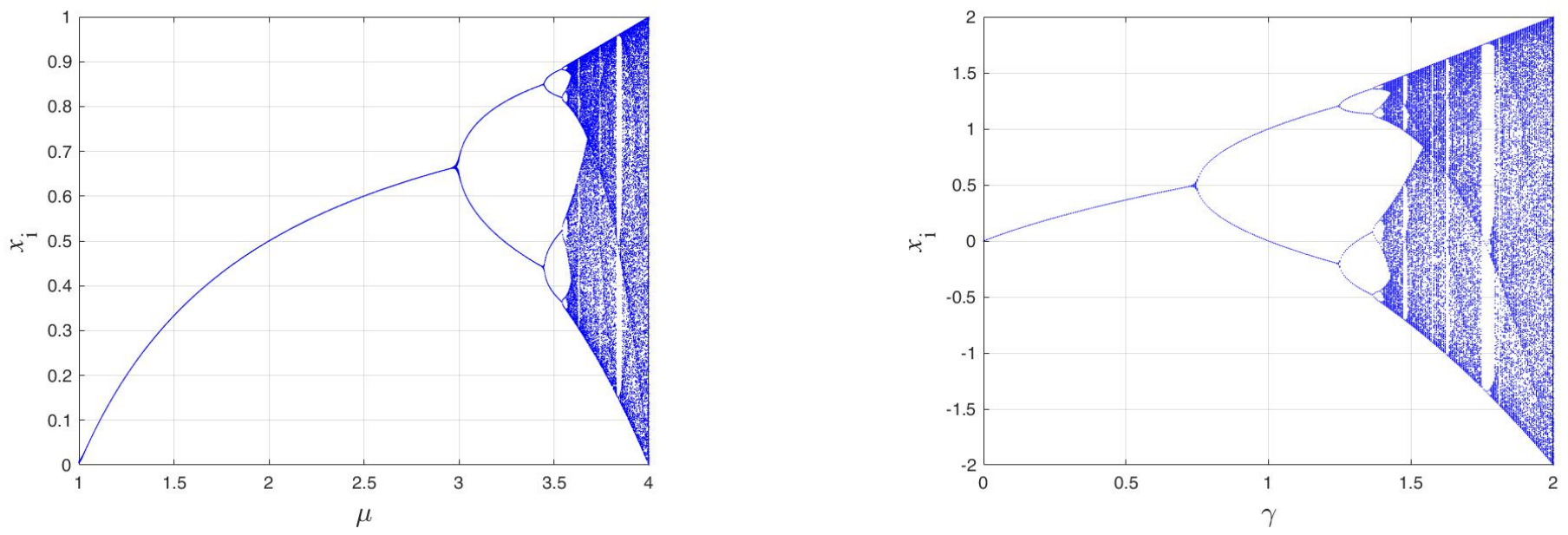

(a) Logistic map (b) Quadratic map

Fig. 3 Bifurcation diagrams of seed maps

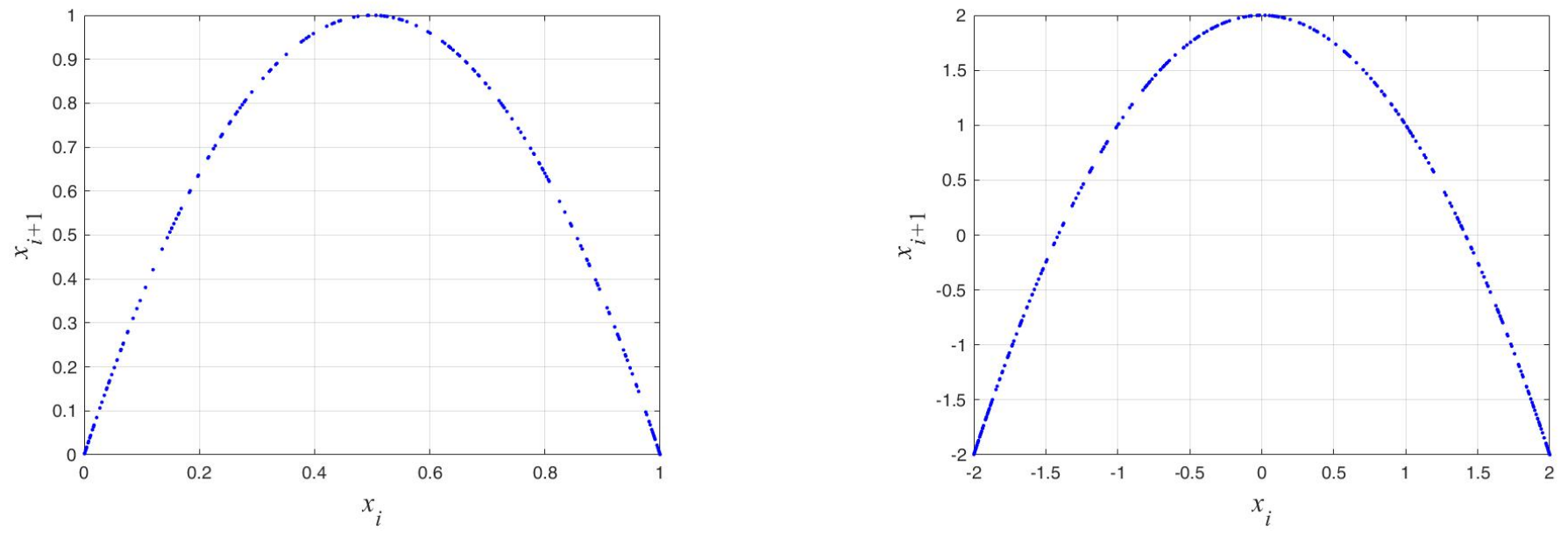

(a) Logistic map ( $\mu$=4 ) (b) Quadratic map ( $\gamma$=2 )

Fig. 4 Phase diagrams of seed maps

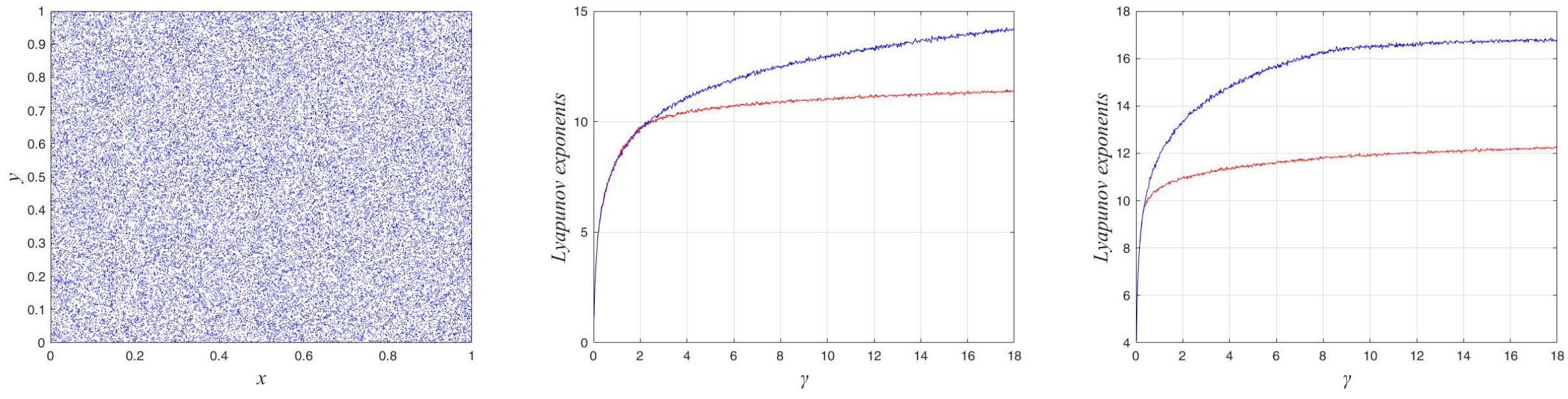

(a) Phase diagram($k$=5) (b) Lyapunov exponents($k$=5) (c) Lyapunov exponents($k$=7)

Fig. 5 Phase diagram and Lyapunov exponents of 2D-ECM

## 4.2 Construction of a 2D-ECM

To counteract the problems of lacking of ergodicity and short iteration period existed in 1D chaotic maps, we need to construct a 2D-ECM [17] with ergodicity based on seed maps, which can be expressed as Eq. (7).

$$\begin{cases} x(i+1) = 2^k \gamma (x(i) + y^2(i)) \bmod 1 \\ y(i+1) = 3^k \gamma (y(i) - x^2(i+1)) \bmod 1 \end{cases}, \quad (7)$$

where state variables $x(0), y(0) \in (0,1)$, control parameter $\gamma \in (0,18]$ are double precision floating point numbers, and exponent $k \in [3,17]$ is an integer. From Fig. 5(a) we can infer that the 2D-ECM has ergodicity in its phase space, and from the Lyapunov exponent spectrum with $k$=5 and $k$=7 shown in Fig. 5(b) and Fig. 5(c), we can infer that the 2D-ECM is non-degenerate, and with the increase of parameter $\gamma$, the 2D-ECM exhibits gradually enhanced hyper chaotic behavior.

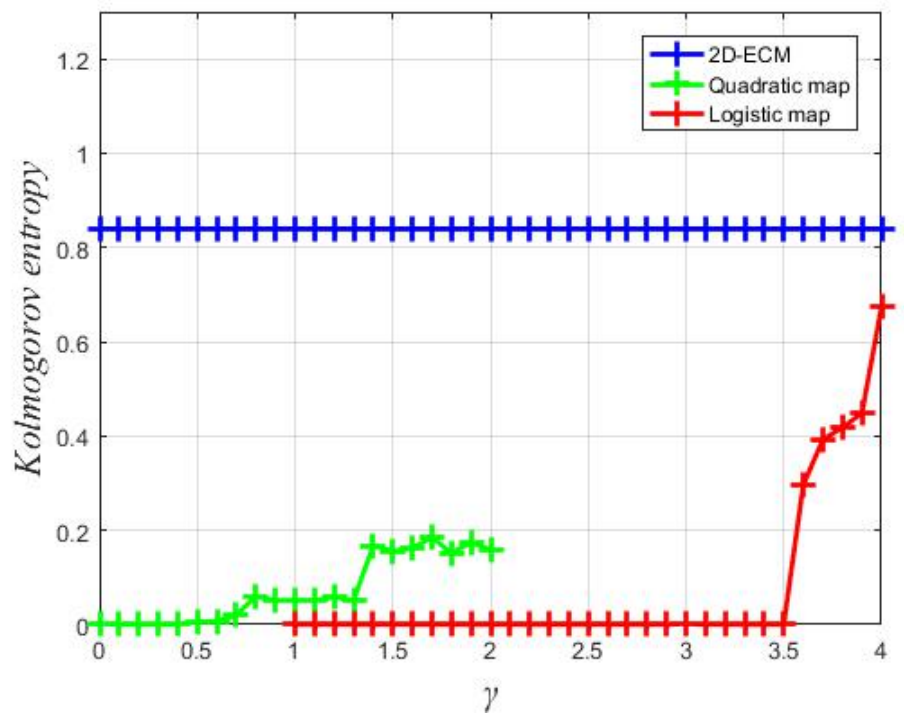

Fig. 6 KE comparison of 3D-ECM with seed maps

## 4.3 Kolmogorov entropy (KE)

Recently, some prediction methods for chaotic time series have been proposed [18], the evolution trajectory of a chaotic map can be predictable in the short-term, and it is impossible to give an accurate, long-term prediction [19]. For any chaotic map, a larger KE [20] indicates that it has a more unpredictable trajectory.

The KEs of 2D-ECM and seed maps with varying control parameter are shown in Fig. 6, which demonstrates that the chaotic time series generated by 2D-ECM has much better unpredictability.

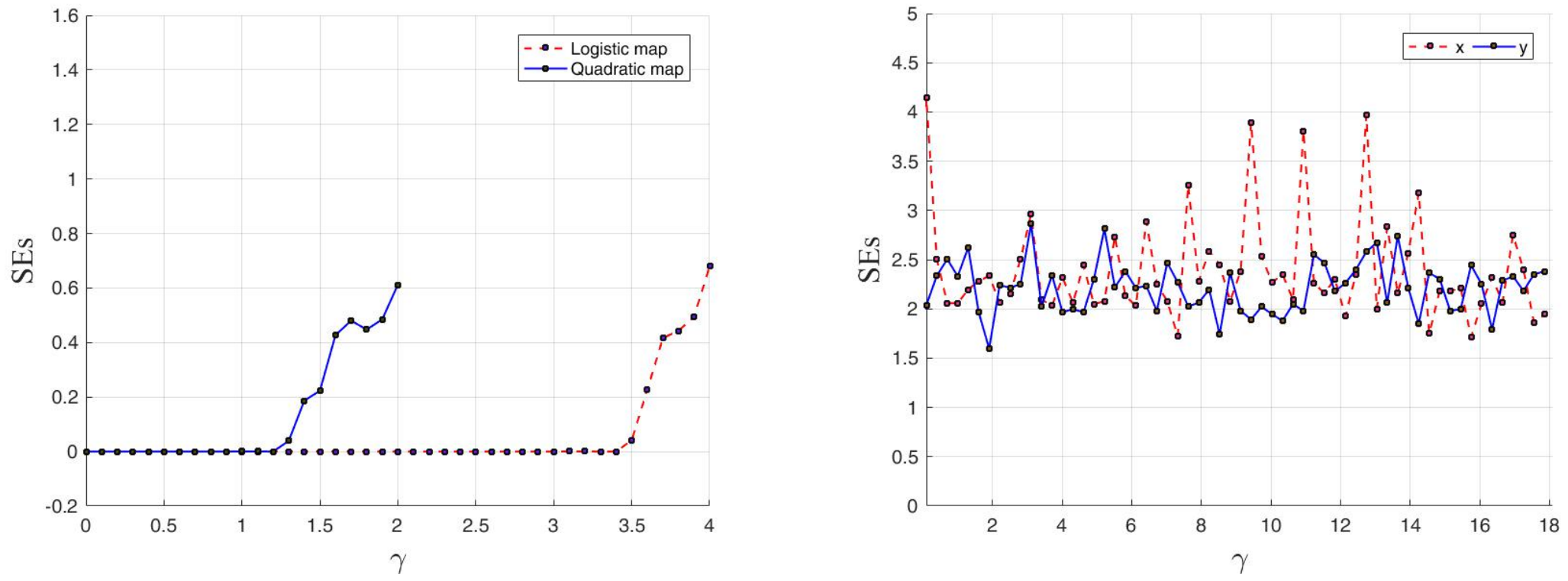

(a) SEs of seed maps (b) SEs of 2D_ECM

Fig. 7 SE comparison between 2D-ECM and seed maps

## 4.4 Sample entropy

The sample entropy(SE) [21] can quantitatively describe the similarity of a time series generated by any chaotic map. A larger SE means a lower degree of regularity, i.e., higher complexity.

The SE of a time series $\{x_1, x_2, ..., x_n\}$ with a given dimension $m$ can be expressed using Eq. (8).

$$SE(m,r,n) = -\log\frac{A}{B}, \tag{8}$$

where dimension $m$ and distance $r$ are usually set to be 2 and $0.2 \times std$, where $std$ denotes the standard deviation of a time series, $A$ and $B$ denote the numbers of vectors $d[X_{m+1}(i), X_{m+1}(j)] < r$ and $d[X_m(i), X_m(j)] < r$, here the template vector $X_m(i) = \{x_i, x_{i+1}, ..., x_{i+m-1}\}$, and $d[X_m(i), X_m(j)]$ denotes the Chebyshev distance between $X_m(i)$ and $X_m(j)$.

From Fig. 7 we can infer that, compare with two seed maps, the chaotic time sequences generated from 2D-ECM have significantly larger SEs, which indicates that it has more complex trajectory and can be used generate more random discrete chaotic time sequences.

## 4.5 Correlation dimension (CD)

As one type of fractal dimension, CD can be applied to measure the dimensionality of the space occupied by a set of discrete state points[22]. Here we calculate the CD values of state point sequences generated from 2D-ECM, it is expected that a $n$-dimension chaotic map should have $n$ CD values around $n$. From the estimation results shown in Fig. 8, we can find that the CD values are all close to 2, which means that the proposed 2D-ECM has better chaotic complexity.

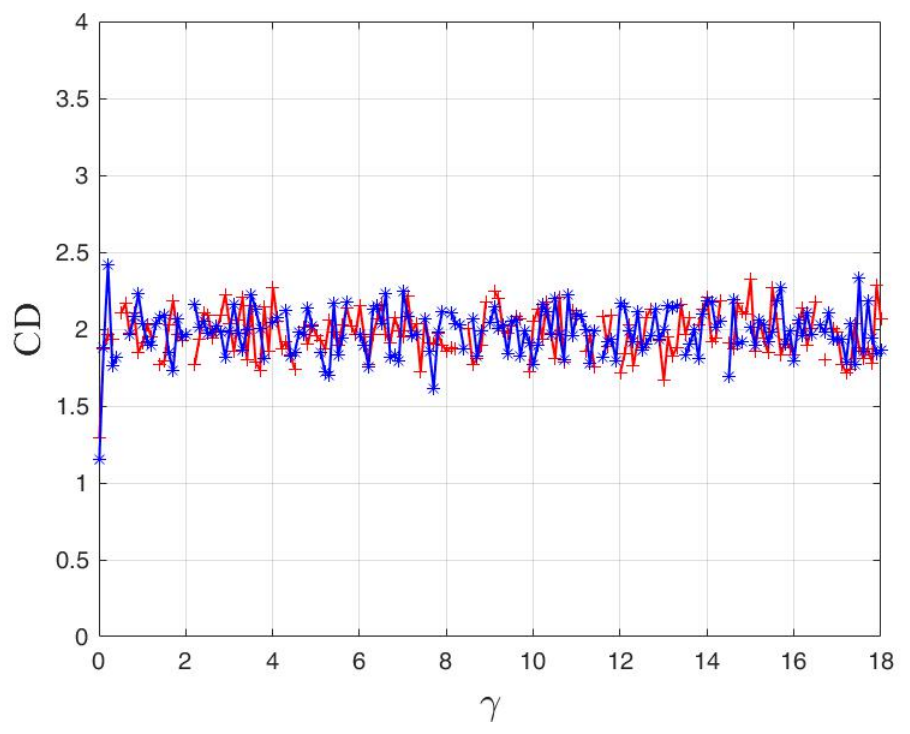

Fig. 8 Correlation dimension of 2D-ECM

Table 3 TestU01 testing comparison

| Gain | | $10^{13}$ | $10^{14}$ | $10^{15}$ | $10^{16}$ |
|---|---|---|---|---|---|
| Quadratic Map [24] | Small Crush | 6 | 4 | 4 | 4 |
| | Crush | 107 | 114 | 102 | 101 |
| Logistic map [24] | Small Crush | 10 | 7 | 3 | 1 |
| | Crush | 104 | 98 | 88 | 75 |
| 2D-ECM($x$, $y$) | Small Crush | (0, 0) | (0, 0) | (0, 0) | (0, 0) |
| | Crush | (3, 2) | (1, 0) | (0, 1) | (0, 0) |
| | Big Crush | (11, 8) | (0, 1) | (0, 0) | (0, 0) |

## 4.6 Randomness testing

Here we use TestU01 [23] to test and compare the randomness of two time series generated by 2D-ECM with seed maps, and set the exponent $k$=7 and the gain $\alpha$=$10^m$, $m$ =13, 14, 15 and 16 respectively. From Table 3 we can infer that the randomness of $(x, y, z)$ is greatly enhanced in comparison with the seed maps.

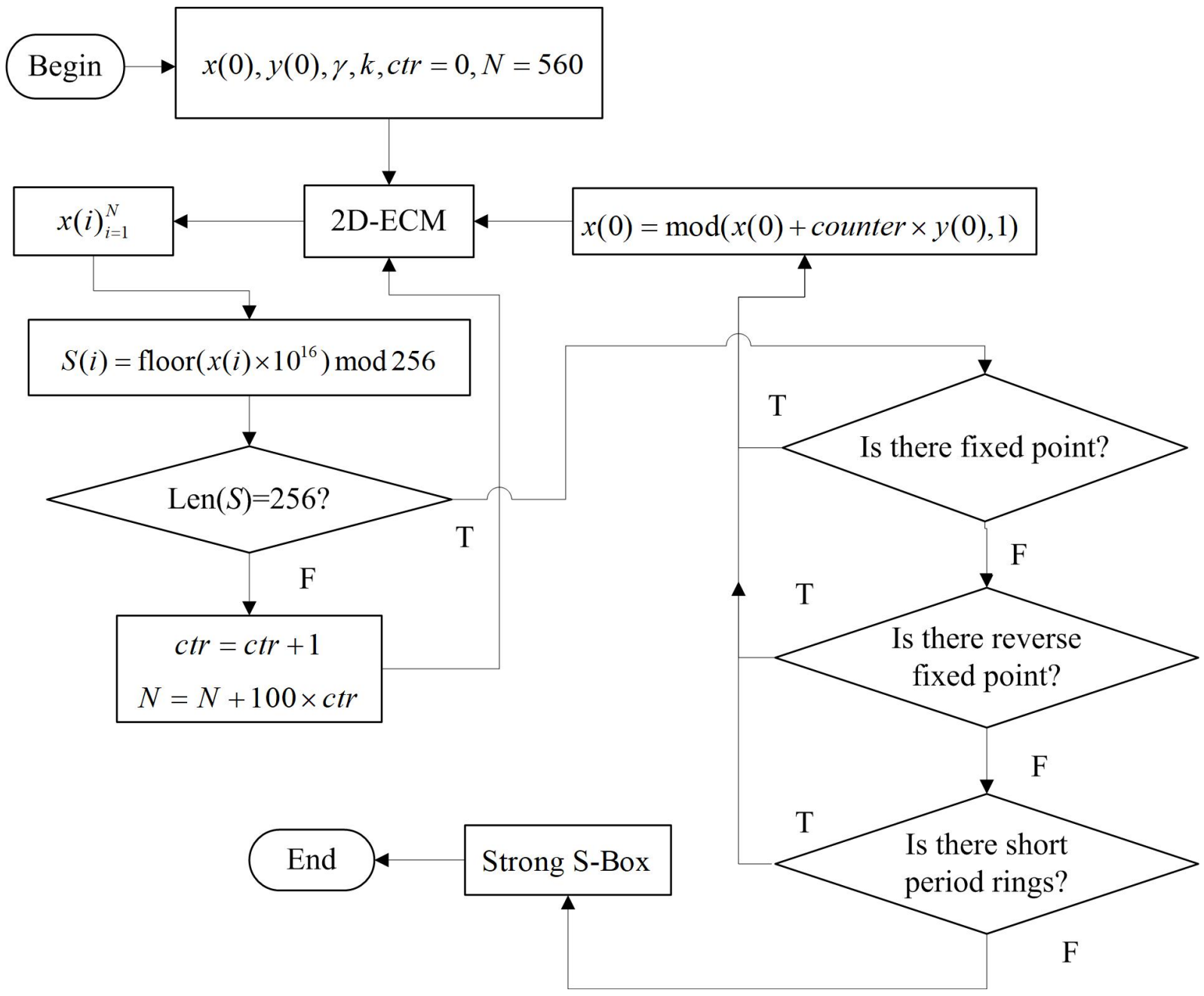

Fig. 9 Flowchart of constructing a strong S-Box

# 5. Construction and analysis of keyed strong S-Box

## 5.1 Construction of keyed strong S-Box

In Ref. [2], we have designed a strong S-Box construction algorithm that satisfies six criteria and conditions (i) and (ii), here we can further construct a strong S-Box based on 2D-ECM and backtracking that satisfies six criteria and conditions (i), (ii) and (iii).

Input: initial conditions $x(0)$, $y(0)$, $\gamma$ and $k$, a counter $ctr = 0$.

Output: A strong 8×8 S-Box.

Step 1. Iterating Eq. (3) 300 times to remove the transient process, then continue to iterate it $N$ times with initial conditions, generally $N \geq 560$, select one of its state points sequences, such as $x(i)|_{i=1}^{N}$, and transform it into an integer sequence $S \in [0, 255]$ using Eq. (9).

$$S(i) = \text{floor}(x(i) \times 10^{16}) \bmod 256. \tag{9}$$

Step 2. Remove the duplicate values in $S$, and retain the first one, if the length of $S$ equals to 256, then go to Step 3, or we can set $ctr = ctr + 1$, $N = N + 100 \times ctr$ and got Step 1, until a sequence $S$ with 256 different integers is obtained.

Step 3. Further check if there exists fixed-point, reverse fixed-point, or short period ring in $S$. If the result is true, then we can modify one of the initial values using Eq. (10), to obtain an updated $x(0)$ and go to Step 1, or got Step 4.

$$x(0) = (x(0) + ctr \times y(0)) \bmod 1, \tag{10}$$

Step 4. Reshape and transform $S$ into a hexadecimal 2D vector $S_B \in [00, \text{FF}]$, then a strong S-Box is constructed. The flowchart of constructing a strong S-Box is shown in Fig. 9.

## 5.2 Analysis of keyed strong S-Box

The strong S-Box based on initial values $x(0)$ =0.414213562373095, $y(0)$ =0.732050807568877, $\gamma$ =5.385164807134504, and $k$=7 is shown in Table 4. Compare with the number of S-Boxes that satisfy three conditions, the ideal number of S-Boxes generated by the proposed algorithm is only $10^{48}$, hence, it is so sparse.

Table 4 A strong S-Box without fixed point, reserve fixed point or short period rings

| | 0 | 1 | 2 | 3 | 4 | 5 | 6 | 7 | 8 | 9 | A | B | C | D | E | F |
|---|---|---|---|---|---|---|---|---|---|---|---|---|---|---|---|---|
| 0 | 2F | BC | 49 | EB | 21 | 73 | 30 | 47 | C6 | 0F | 0C | 3E | 9C | 3F | 44 | 75 |
| 1 | 14 | B0 | B4 | 4E | C9 | E8 | 2C | C8 | D5 | 0D | F8 | 82 | 0A | FE | 5F | 34 |
| 2 | 04 | 42 | E9 | EA | B8 | 4C | 6A | 56 | 58 | C1 | FF | DC | E1 | 9E | B7 | 6E |
| 3 | A6 | 13 | AA | D0 | 1A | 29 | 0E | 57 | F2 | 09 | 68 | 11 | 05 | 3B | 62 | 7C |
| 4 | 3D | 1D | 90 | F7 | E0 | 4F | 15 | 20 | A7 | F5 | B6 | 4A | 27 | 25 | 35 | C4 |
| 5 | 52 | 6C | 2E | 26 | 9B | 79 | C3 | 92 | A3 | 53 | 02 | 88 | 48 | 1C | 22 | 28 |
| 6 | 51 | 4B | 87 | 40 | 63 | E4 | 1E | DF | FA | 80 | DD | D9 | 83 | 23 | 50 | 6B |
| 7 | 8A | A9 | 41 | 08 | 97 | 07 | BA | DA | 76 | A4 | A5 | B2 | D2 | 8E | D4 | AF |
| 8 | D7 | F3 | 84 | 03 | E6 | B1 | D8 | 01 | D3 | 32 | 86 | 8F | 38 | BE | CA | C0 |
| 9 | 2A | 71 | 59 | 10 | 3C | 64 | F0 | 98 | AE | 9A | 70 | 9F | 18 | F9 | 7A | 96 |
| A | B9 | CC | 67 | 6F | C2 | 39 | E2 | 8B | EC | C7 | 19 | E5 | 5D | CE | 4D | 6D |
| B | 78 | 81 | AC | 3A | A2 | 0B | 36 | DB | 5E | 37 | 7E | 8D | 55 | A0 | AD | B5 |
| C | 77 | 5C | 89 | D1 | 72 | E3 | 7B | F1 | 06 | 7F | A8 | CF | 1F | FD | 5B | 1B |
| D | 95 | 99 | 16 | 33 | FC | EF | 94 | 45 | E7 | 17 | BD | BB | 7D | D6 | 2D | ED |
| E | 00 | A1 | 5A | F4 | 9D | 31 | BF | 60 | CD | 91 | 8C | 61 | F6 | C5 | 24 | 93 |
| F | 43 | 65 | 54 | 46 | 85 | 74 | DE | 69 | B3 | 2B | 66 | 12 | FB | AB | EE | CB |

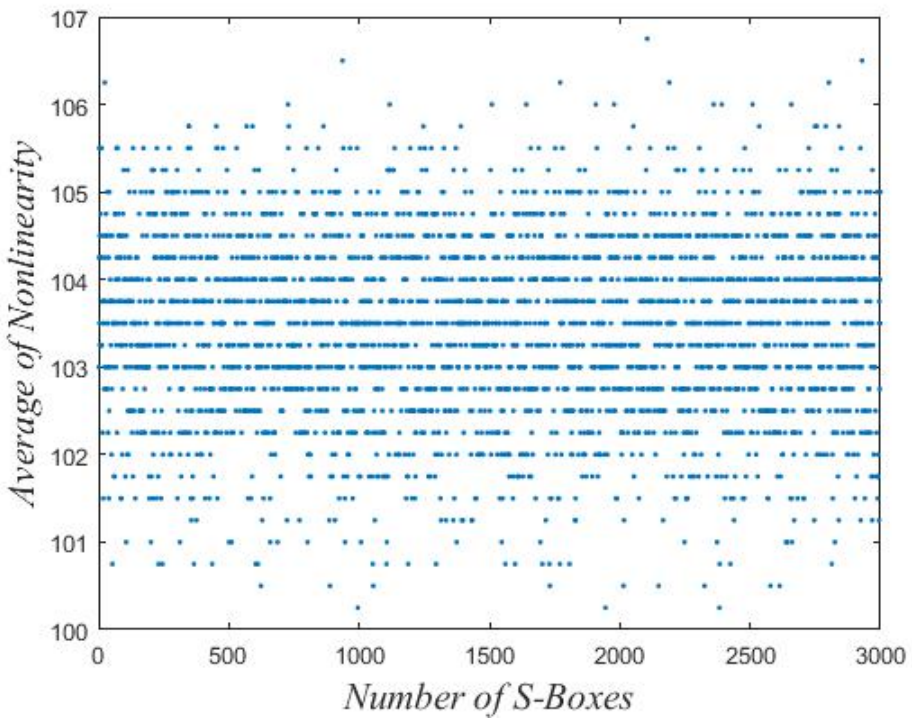

Fig. 10 Distribution of average nonlinearity of 3000 S-Boxes

Here we set random initial values of 2D-ECM, to generate 3000 strong S-Boxes, their averages of nonlinearity are shown in Fig. 10, and the performance comparison of six criteria with other S-Boxes are shown in Table 5, from which we can find that, although the S-Boxes of AES, SM4 and ZUC have high nonlinearity, they still have weaknesses. Here we adjust them to satisfy conditions (i), (ii) and (iii) and re-test their performance, from Table 5 we can infer that, the six criteria and three conditions are interinhibitive.

Table 5 Performance comparison of S-Boxes

| S-Box | Nonlinearity | | | SAC | | | BIC-SAC | BIC-Nonlinearity | DAP | LAP |
|---|---|---|---|---|---|---|---|---|---|---|
| | Min | Max | Avg. | Min | Max | Avg. | | | | |
| Table 4 | 100 | 108 | 103.50 | 0.3750 | 0.5947 | 0.4980 | 0.5024 | 103.20 | 0.0391 | 0.1406 |
| 3000 strong S-Box | - | - | 103.45 | - | - | 0.5009 | 0.5019 | 103.45 | 0.0441 | 0.1381 |
| S-Box in AES | 112 | 112 | 112.00 | 0.4531 | 0.5625 | 0.5049 | 0.5046 | 112.00 | 0.0156 | 0.0625 |
| Adjusted S-Box in AES | 110 | 112 | 111.25 | 0.4375 | 0.5625 | 0.5049 | 0.5051 | 111.21 | 0.0234 | 0.0781 |
| S-Box in SM4 | 112 | 112 | 112.00 | 0.4375 | 0.5625 | 0.4997 | 0.5049 | 112.00 | 0.0156 | 0.0625 |
| Adjusted S-Box in SM4 | 110 | 112 | 110.25 | 0.4375 | 0.5781 | 0.4997 | 0.5052 | 109.86 | 0.0234 | 0.0859 |
| $S_0$ in ZUC | 96 | 104 | 98.00 | 0.3437 | 0.6250 | 0.4949 | 0.4951 | 101.70 | 0.0312 | 0.1250 |
| Adjusted $S_0$ in ZUC | 94 | 102 | 97.25 | 0.3437 | 0.6250 | 0.4951 | 0.4952 | 101.07 | 0.0391 | 0.1328 |
| $S_1$ in ZUC | 112 | 112 | 112.00 | 0.4375 | 0.5625 | 0.5093 | 0.5058 | 112.00 | 0.0156 | 0.0625 |
| Adjusted $S_1$ in ZUC | 112 | 110 | 111.25 | 0.4375 | 0.5781 | 0.5095 | 0.5053 | 110.93 | 0.0234 | 0.0781 |
| S-Box in Whirlpool | 100 | 107 | 103.38 | 0.3672 | 0.6406 | 0.5073 | 0.4974 | 104.46 | 0.0313 | 0.1250 |
| Adjusted S-Box in Whirlpool | 100 | 107 | 104.13 | 0.3750 | 0.6250 | 0.5068 | 0.4980 | 104.46 | 0.0313 | 0.1250 |
| S-Box in Skipjack | 103 | 108 | 105.63 | 0.3906 | 0.5938 | 0.5032 | 0.4997 | 104.04 | 0.0469 | 0.1133 |
| Adjusted S-Box in Skipjack | 103 | 108 | 106.13 | 0.3906 | 0.5938 | 0.5034 | 0.4994 | 104.04 | 0.0469 | 0.1133 |
| S-Box in Fig. 9 [11] | 114 | 116 | 114.5 | 0.4531 | 0.5938 | 0.5012 | 0.5047 | 104.21 | 0.0391 | 0.1406 |
| Adjusted S-Box in Fig. 9 [11] | 110 | 114 | 112.5 | 0.4375 | 0.5781 | 0.5015 | 0.5043 | 103.86 | 0.0391 | 0.1484 |

# 6. Key expansion algorithm and evaluation

Benefiting from the irreversibility of the chaotic iteration process, we can design an irreversible key expansion algorithm with the following characteristics: (1) the initial key can not be recovered from any round key, and (2) the round keys are independent of each other.

Generally, the 256-bit initial key $IK$, can be represented as $IK = (ik_1, ik_2, ik_3, ..., ik_n)$, where $n$ =32 , and $ik_i$ represents an 8-bit byte, $IK$ can be transformed into the initial values of 2D-ECM, or served as impulse to perturb the 2D-ECM [25]. Similarly, the *r*-th round key can be represented as $rK_r = (rk_1, rk_2, rk_3, ..., rk_n)$.

## 6.1 Algorithm description

Input: The 256-bit initial key *IK*, round number $r$ .

Output: round keys $rK$.

Step 1. Calculate the 256-bit Hash value $H_b(IK)$ of $IK$ using SHA2-256, to eliminate the deviation of 0 and 1, then express it as a hexadecimal number $H_h(IK)$ with the size of $32 \times 2$. Substitute each pair of hexadecimal numbers with a strong S-Box, to get $H_{hS}$.

Step 2. Transform $H_{hS}$ into the initial values of 2D-ECM using Eq. (11) and Eq. (12), to get the initial values $x(0)$ and $y(0)$ of 2D-ECM.

$$\begin{cases} x_t(1:8) = \text{bitxor}(\text{hex2dec}(H_{hS}(1:8)), \text{hex2dec}(H_{hS}(9:16))) \\ y_t(1:8) = \text{bitxor}(\text{hex2dec}(H_{hS}(17:24)), \text{hex2dec}(H_{hS}(25:32))) \end{cases}, \tag{11}$$

$$\begin{cases} x(0) = x_t(1)\times 10^{-3} + x_t(2)\times 10^{-6} + x_t(3)\times 10^{-9} + x_t(4)\times 10^{-10} + \\ \qquad x_t(5)\times 10^{-11} + x_t(6)\times 10^{-12} + x_t(7)\times 10^{-13} + x_t(8)\times 10^{-14} \\ y(0) = y_t(1)\times 10^{-3} + y_t(2)\times 10^{-6} + y_t(3)\times 10^{-9} + y_t(4)\times 10^{-10} + \\ \qquad y_t(5)\times 10^{-11} + y_t(6)\times 10^{-12} + y_t(7)\times 10^{-13} + y_t(8)\times 10^{-14} \end{cases}. \tag{12}$$

Step 3. Iterate Eq. (3) 300 times with $x(0)$ and $y(0)$ to remove the influence of transient process, continue to iterate it $4r$ times, to get $x(1:4r)$ and $y(1:4r)$, then transform them into round keys using Eq. (13), to get $rk_x(i)$ and $rk_y(i)$ with 8 hexadecimal numbers.

$$\begin{cases} rk_x(i) = \text{substr}(\text{dec2hex}(10^{14}+\text{floor}(x(i)\times 10^{14})), 2:9) \\ rk_y(i) = \text{substr}(\text{dec2hex}(10^{14}+\text{floor}(y(i)\times 10^{14})), 2:9) \end{cases}, \quad i = 1, 2, \ldots, 4r, \tag{13}$$

where function substr(*s*, *start*: *end*) is to truncate a sub-string from the starting position to the ending position of *s*.

Step 4. Finally, combine $rk_x$ and $rk_y$ into $rK(j)$ to serve as the round keys using Eq. (14).

$$rK(j) = rk_x(4(j-1)+1:4j) \,||\, rk_y(4(j-1)+1:4j), \quad j = 1, 2, \ldots, r. \tag{14}$$

Table 6 The 256-bit round keys and Hamming distance

| Round | Initial key and round key | Hamming distance |
|---|---|---|
| *IK* | 0000000000000000000000000000000000000000000000000000000000000000 | |
| 1 | 9A0D0E3E031E7BB0345A43A5EB9895BC672CF98D46B6399E8C5E238E081A864B | 123 |
| 2 | 7AA2165A25E2ACE14500650C6F5ECFB3CF6B4ECECEBF9BF28F78536DBECDA25A | 139 |
| 3 | 8A9D00D19CB61F6CF0D5E5458647B18F7473E83469F55EE87EDD40AE9C0A1293 | 128 |
| 4 | 84D0E4493C6D6BB09FBB91BAE40886D7D1789A3B7A5B65FC0B83892794207687 | 126 |
| 5 | B4D5CCDBDFE0C55900A8D7F1D9B516CFDF531A5178CF2EC5C2772143F66DFE7D | 143 |
| 6 | 2DE27A1F19D2AD352C25D7453D48682A584F8787E99DFE5F2CDE02666260D31A | 128 |
| 7 | 814171D38E9D183DF4F95B453DB402F58FD10B4D0AB38CB707FB6590E728AE0B | 128 |
| 8 | 6BDDCB80C1D82CB4BFF419BAAC5E10DE645F8385E38CC9B89BDE6BC3DA7A1146 | 133 |
| 9 | 9A0D0E3E031E7BB0345A43A5EB9895BC672CF98D46B6399E8C5E238E081A864B | 123 |
| 10 | 7AA2165A25E2ACE14500650C6F5ECFB3CF6B4ECECEBF9BF28F78536DBECDA25A | 139 |
| 11 | 8A9D00D19CB61F6CF0D5E5458647B18F7473E83469F55EE87EDD40AE9C0A1293 | 128 |
| 12 | 84D0E4493C6D6BB09FBB91BAE40886D7D1789A3B7A5B65FC0B83892794207687 | 126 |
| 13 | B4D5CCDBDFE0C55900A8D7F1D9B516CFDF531A5178CF2EC5C2772143F66DFE7D | 143 |
| 14 | 2DE27A1F19D2AD352C25D7453D48682A584F8787E99DFE5F2CDE02666260D31A | 128 |
| 15 | 814171D38E9D183DF4F95B453DB402F58FD10B4D0AB38CB707FB6590E728AE0B | 128 |
| 16 | 6BDDCB80C1D82CB4BFF419BAAC5E10DE645F8385E38CC9B89BDE6BC3DA7A1146 | 133 |

### 6.2 Hamming distance

The round keys generated by the key expansion algorithm are independent of each other. The corresponding round keys with Hamming distance to initial key are given in Table 6. For 1000 round keys, the average value of Hamming distance is 128.0132, which is very close to the ideal value 128.

### 6.3 Flexibility

The proposed key expansion algorithm is flexible: (1) by setting a different parameter $n$, it can produce 256, 512, 1024-bit or longer round key, (2) it can generate more round keys by setting the parameter $r$, and (3) the S-Box can be designed to be keyed or un-keyed, and the former can generate more secure round keys.

### 6.4 Security analysis

Benefiting from the one-way character of chaotic iteration and information redundancy, the proposed key expansion algorithm is irreversible: (1) the initial key can not be transformed from any round key, and (2) the current round key can not be transformed from its previous or next round key. For example, to produce each round key with 64 hexadecimal numbers (256 bit), we can iterate the 2D-ECM 4 times, to get 8 pairs of double precision state points, transform each pair of them into 12 hexadecimal numbers, then truncate 8 numbers from them, to form 64 hexadecimal numbers as a round key.

## 7. Conclusion

The cryptanalysis results of S-Boxes revealed some weaknesses inside, such as fixed-point, reverse fixed-point, and short periodic rings. To eliminate the weaknesses, we calculated the number of S-Boxes that satisfies three conditions, and constructed a 2D-ECM with ergodicity in a larger parameter range, and based on it, designed a keyed strong S-Box construction algorithm, and based on it, we designed a key expansion algorithm without reversibility, to solve the reversibility of key expansion in block cipher, to make the round keys independent of each other. Experimental results evaluation demonstrated the effectiveness of key expansion.

### Acknowledgments

This research is supported by the National Natural Science Foundation of China (No: 61662073), the Science and Technology Program of University of Jinan (No: XKY2070).

### References

[1] Al-Muhammed, Muhammed. A novel key expansion technique using diffusion[J]. Computer Fraud & Security, 2018, 2018(3): 12-20.

[2] Liu H, Kadir A, Xu C. Cryptanalysis and constructing S-Box based on chaotic map and backtracking[J]. Applied Mathematics and Computation, 2020, 376: 125153.

[3] Daemen, J. and Rijmen, V. The Design of Rijndael: AES—The Advanced Encryption Standard of Information Security and Cryptography. Springer Verlag, Berlin, 2002, 22: 231-240.

[4] State Cryptography Administration Office of Security Commercial Code Administration. SM4 block cipher algorithm: GM/T 0002-2012[S]. Beijing: China Standard Press, 2012(in Chinese).

[5] Li C, Feng B, et al. Dynamic analysis of digital chaotic maps via state-mapping networks. IEEE Transactions on Circuits and Systems I: Regular Papers, 2019, 66(6): 2322-2335.

[6] Detombe J, Tavares S. Constructing large cryptographically strong S-Boxes[C]//International Workshop on the Theory and Application of Cryptographic Techniques. Springer, Berlin, Heidelberg, 1992: 165-181.

[7] Wang Y, Zhang Z, et al. A genetic algorithm for constructing bijective substitution boxes with high nonlinearity[J]. Information Sciences, 2020, 523: 152-166.

[8] Picek S, Santana R, Jakobovic D. Maximal nonlinearity in balanced boolean functions with even number of inputs, revisited. 2016 IEEE Congress on Evolutionary Computation (CEC), 3222-3229. DOI: 10.1109/CEC.2016.7744197.

[9] Isa H, Jamil N, et al. Construction of Cryptographically Strong S-Boxes Inspired by Bee Waggle Dance[J]. New Generation Computing, 2016, 34: 221-238.

[10] Dragan L. A novel method of S-Box design based on chaotic map and composition method[J]. Chaos, Solitons Fractals, 2014, 58: 16-21.

[11] Miroslav M. On the design of chaos-based S-boxes[J]. IEEE Access, 2020, DOI: 10.1109/ACCESS.2020.3004526.

[12] Hua Z, Li J, Chen Y, et al. Design and application of an S-Box using complete Latin square [J]. Nonlinear Dynamics, 2021, 104: 807-825.

[13] Ibrahim S, Abbas A M. Efficient Key-dependent Dynamic S-Boxes Based on Permutated Elliptic Curves[J]. Information Sciences, 2021, 558: 246-264.

[14] Doerrie H . 100 great problems of elementary mathematics. Their history and solution. Dover Publications, 1965.

[15] May, Robert M. Simple mathematical models with very complicated dynamics[J]. 1976, 261(5560): 459-467.

[16] Jakobson M V. Absolutely continuous invariant measures for one-parameter families of one-dimensional maps[J]. Communications in Mathematical Physics, 1981, 81(1): 39-88.

[17] Liu H, Kadir A, Xu C. Color image encryption with cipher feedback and coupling chaotic map. International Journal of Bifurcation and Chaos. 2020, 30(12): 2050173.

[18] Na X, Ren W, Xu X. Hierarchical delay-memory echo state network: A model designed for multi-step chaotic time series prediction[J]. Engineering Applications of Artificial Intelligence, 2021, 102(1):104229.

[19] Tang L, Bai Y, et al. A hybrid prediction method based on empirical mode decomposition and multiple model fusion for chaotic time series. Chaos, Solitons & Fractals, 2020, 141: 110366.

[20] Paulo R, M Baptista, Isabel S. Density of first Poincaré returns, periodic orbits, and Kolmogorov–Sinai entropy[J]. Communications in Nonlinear Science & Numerical Simulation. 2011, 16(2): 863-875.

[21] Richman, Joshua, S, et al. Physiological time-series analysis using approximate entropy and sample entropy[J]. American Journal of Physiology Heart & Circulatory Physiology, 2000, 278(6): H2039-H2049.

[22] Theiler, James. Efficient algorithm for estimating the correlation dimension from a set of discrete points[J]. Physical Review A, 1987, 36(9): 4456-4462.

[23] L'Ecuyer P, Simard R. TestU01: A C library for empirical testing of random number generators[J]. ACM Transactions on Mathematical Software, 2007, 33(4, article 22).

[24] Liu H, Wang X, Kadir A. Constructing chaos-based hash function via parallel impulse perturbation[J]. Soft Computing, 2021: 2021, 25(16): 11077-11086

[25] Liu H, Zhang Y, Kadir A, et al. Image encryption using complex hyper chaotic system by injecting impulse into parameters[J]. Applied Mathematics and Computation, 2019, 360: 83-93.